\documentclass[preprint,12pt]{elsarticle}
\usepackage{amssymb}
\usepackage{placeins}
\usepackage{rotating}
\usepackage{colortbl}
\usepackage{lscape}
\usepackage{longtable}
\usepackage{adjustbox}
\usepackage{pdflscape}
\newcolumntype{L}{>{\centering\arraybackslash}m{6cm}}
\newcolumntype{M}{>{\centering\arraybackslash}m{2.5cm}l}

\biboptions{sort&compress}

\journal{Artificial Intelligence in Medicine}

\begin{document}

\begin{frontmatter}

\title{The Effect of Acute Stress on the Interpretability and Generalization of Schizophrenia Predictive Machine Learning Models}
\author[inst1]{Gideon Vos}

\affiliation[inst1]{organization={College of Science and Engineering, James Cook University},
            addressline={James Cook Dr}, 
            city={Townsville},
            postcode={4811}, 
            state={QLD},
            country={Australia}}
            
\author[inst1]{Maryam Ebrahimpour}
\author[inst2]{Liza van Eijk}
\author[inst3]{Zoltan Sarnyai}
\author[inst1]{Mostafa Rahimi Azghadi}

\affiliation[inst2]{organization={College of Health Care Sciences, James Cook University},
            addressline={James Cook Dr}, 
            city={Townsville},
            postcode={4811}, 
            state={QLD},
            country={Australia}}

\affiliation[inst3]{organization={College of Public Health, Medical, and Vet Sciences, James Cook University},
            addressline={James Cook Dr}, 
            city={Townsville},
            postcode={4811}, 
            state={QLD},
            country={Australia}}
\begin{abstract}
\paragraph{Introduction}
\noindent Schizophrenia is a chronic and severe mental disorder and early diagnosis is crucial to improve patient outcomes. Despite decades of ongoing research, the complexity and heterogeneity of schizophrenia make it difficult to predict its onset and progression. Electroencephalography (EEG) has emerged as a promising tool for exploring the neural underpinnings of schizophrenia, and machine learning (ML) approaches have been increasingly applied to EEG data to support schizophrenia diagnosis. Mental health conditions often overlap, thereby complicating the interpretability of ML-based predictive models which is essential for ensuring the reliability and clinical applicability of predictive models in schizophrenia research. In this paper, we first assess the accuracy of predictive machine learning models developed using open EEG datasets to predict schizophrenia. We then assess the potential impact of subjects' stress during EEG recording on the predictive performance of these models. Finally, our analysis incorporates acute stress prediction into the pipeline to qualify the EEG-based training data, demonstrating that overlapping health conditions, such as acute stress during EEG recording, can adversely impact model predictive performance.

\paragraph{Methods}
\noindent Four open EEG datasets were used for experimentation. These include a large healthy-controls dataset, a dataset labeled for subjects experiencing acute stress, a dataset of schizophrenia patients recorded at rest, and a dataset of schizophrenia patients recorded while performing a visual discrimination task. Both these schizophrenia datasets included a set of healthy subjects as well. Four machine learning models were built using the XGBoost algorithm, the first to predict acute stress (using the pre-labeled acute stress dataset), the second and third to classify healthy controls from schizophrenia at rest and at task respectively, and the fourth a multi-class model to predict schizophrenia during both at rest or during task from healthy controls. Explainable Artificial Intelligence (XAI) techniques were applied during experimentation and the results were analyzed. Experiments were performed to test the generalization ability of both schizophrenia prediction models with their respective datasets' healthy controls, as well as using external healthy controls that were independently health-screened. The stress prediction model was next applied to identify all subjects showing high acute stress, and these subjects were excluded in subsequent experiments and analyses. Finally, a novel approach was applied to adjust EEG frequency band power as an artifact removal technique in order to compensate for potential acute stress during EEG recordings. The predictive models were re-applied to the original test subjects for analysis and comparison, showing the significant benefits of our approach.

\paragraph{Results}
\noindent Our results show that forms of acute stress vary significantly across EEG recording sessions, affecting model performance and machine learning classification accuracy. Model generalization improved once these varying stress levels were considered and compensated for during model training. The use of a large, validated healthy-control dataset further improved predictive accuracy across experimental models. Our findings highlight the importance of thorough health screening prior to EEG recording, and careful management of the patient's condition during the process. These steps are essential for obtaining high-quality training data for building machine learning models aimed at predicting mental health conditions, particularly due to the potential overlap of various conditions. 

\paragraph{Conclusion}
\noindent Stress and anxiety induced during or by the EEG recording procedure can potentially adversely affect machine learning model generalization. This may require further preprocessing of EEG data by treating acute stress as an additional physiological artifact. Our proposed approach to identify and compensate for stress artifacts in EEG data used for training machine learning models showed a significant improvement in predictive performance. We hope that our method and publicly available code and data could enhance research towards improved patient outcomes.

\end{abstract}




\end{frontmatter}


\section{Introduction}

\noindent Schizophrenia is a chronic and severe mental disorder characterized by distortions in thinking, perception, emotions, language, sense of self, and behavior \cite{DSM2013}. Prior research indicates that stress during adolescence may be a precipitating factor for the onset of schizophrenia, and managing stress during this critical period might prevent the development of psychosis \cite{Gomes2016}. Adolescents with a predisposition to stress hyper-responsivity, or those exposed to significant stressors, can experience a sequence of events leading to a schizophrenia-like profile in adulthood \cite{Gomes2017}. Arnsten \cite{Arnsten2023} noted that stress exacerbates thought disorder in schizophrenia, linked to severe dysfunction in the prefrontal cortex, a region crucial for higher cognitive functions. \\

\noindent Stress further impacts the brain and body differently in individuals with schizophrenia compared to healthy individuals, or those at high risk for psychosis \cite{Schifani2018}. Schifani \emph{et al.} \cite{Schifani2018} found a disrupted stress response in individuals with schizophrenia, a phenomenon not observed in healthy controls or those at high risk for psychosis. Typically, both dopamine and cortisol levels increase during stress in healthy individuals \cite{Joels2009}. However, this correlation was absent in individuals with schizophrenia, indicating a disrupted stress regulatory system. Understanding these differences in stress response may contribute towards developing more effective treatments and diagnosis for schizophrenia.\\

\begin{figure}[h!]
\centering
\includegraphics[width=\textwidth]{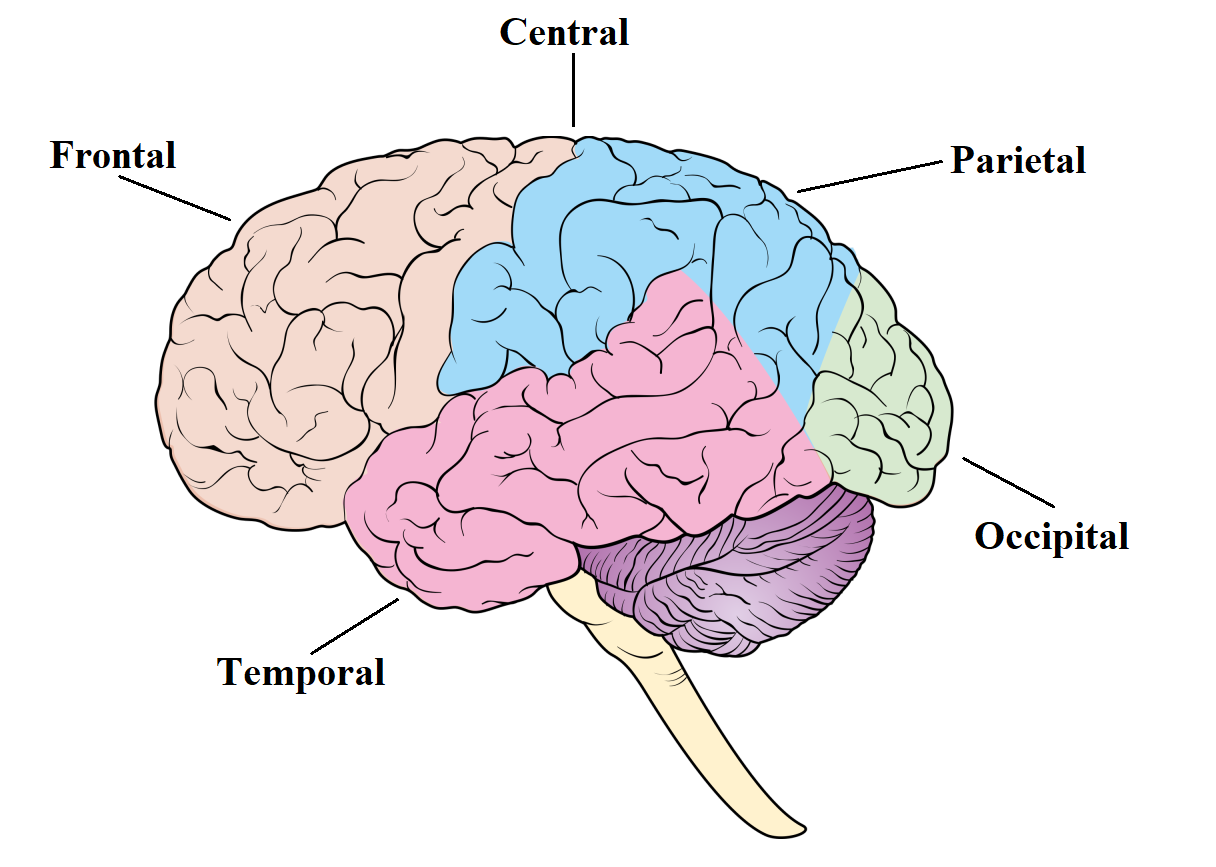}
\caption{\label{fig:figure1}Brain regions commonly used in EEG.}
\end{figure}

\noindent Modern sophisticated technologies such as genomics, neuroimaging, and computational methodologies using machine learning have advanced the study of mental health conditions including schizophrenia \cite{Lieberman2021}, with a number of studies applying machine learning techniques towards classifying schizophrenia from healthy controls using electroencephalogram (EEG) data. Ruiz de Miras \emph{et al.} \cite{RuizdeMiras2023} tested five machine learning algorithms utilizing linear and non-linear measures computed using sliding windows from resting state EEG data labeled for schizophrenia (31 subjects) and healthy controls (20 subjects) training and classifications. They achieved predictive accuracy of 89\% when using Support Vector Machines (SVM), and 87\% accuracy when using Random Forest (RF). Frontal and temporal regions were found to present significant differences between the two categories subjects in a high number of measures. Arias \emph{et al.} \cite{Arias2023} compared SVM, XGBoost (XGB) \cite{XGBoost} and Adaptive Boosting to predict schizophrenia from EEG data. 
Incorporating Explainable A.I. (XAI) techniques, specifically Shapley Additive Explanations (SHAP) \cite{SHAP}, into the pipeline, they achieved 93\% accuracy. SHAP identified the delta band across the temporal and parietal regions (see Figure \ref{fig:figure1}) as the most important features.\\

\FloatBarrier

\noindent A number of further studies have tested the generalization of multi-class models trained to predict a number of different mental health conditions, which included schizophrenia. Emre \emph{et al.} \cite{Emre2023} tested several algorithms using a multi-modal approach, and reported individual at-rest condition accuracy of higher than 89\% when classifying post-traumatic stress disorder, bipolar disorder, depression, schizophrenia, attention deficit hyperactivity disorder and healthy controls, using either the C5.0 or SVM algorithms. Park \emph{et al.} \cite{Park2021} similarly classified multiple trauma and stress-related disorders, addictive disorders, obsessive–compulsive disorders, anxiety disorders, mood disorders, schizophrenia and healthy controls and achieved an area under the curve (AUC) result of 93.83\% ($p<0.001$) for schizophrenia specifically. They noted that the alpha band served as the most important predictive model feature across the central, parietal and occipital regions. \\

\noindent Rahul \emph{et al.} \cite{Rahul2024} performed a systematic review of schizophrenia classification of EEG data using machine learning techniques, and compared both classical algorithms such as SVM and Decision Trees (DT) with results reported by more sophisticated Deep Neural Networks (DNN), reporting higher predictive accuracy on average for deep-learning models compared to classical models. However, the use of model explainability techniques or feature importance findings were not reported, which could be valuable to both clinicians and data scientists in understanding how algorithms classify schizophrenia, and which underlying features are important disease indicators. \\

\noindent Prior studies into the effects of acute stress using EEG reported differences in the delta band to be the most significant compared to non-stressed controls \cite{Halim2020, Pernice2020, Bhatnagar2023}, while alpha was reported by three \cite{Pernice2020, Arsalan2021, Vanhollebeke2022}, beta by three \cite{Arsalan2021, Sharif2023, Vanhollebeke2022}, gamma by two \cite{Halim2020, Parent2020} and theta by a single study \cite{Majid2022}. Frontal and temporal lobes were the most important reported locations for sensor placement \cite{Parent2020, Pernice2020, Bhatnagar2023}. A meta-analysis on stress conducted by Vanhollebeke \emph{et al.} \cite {Vanhollebeke2022} revealed a significant effect size for alpha power, but an insignificant effect size for beta power and frontal alpha asymmetry.\\

\noindent In this study, we aim to enhance the prediction of schizophrenia using machine learning by accounting for acute stress as an underlying factor that may impact model generalization. Towards that end, we initially develop and validate baseline models that distinguish patients with schizophrenia from healthy controls in both a binary classification and multi-class setting. We then integrate XAI techniques into these models and incorporate acute stress measurement into the analytical framework. This examines whether: i) XAI can provide a direct correlation between machine learning model predictions and established mental health research, and ii) XAI can identify underlying health conditions that may impact model accuracy. Leveraging the results of these preliminary experiments, we demonstrate that acute stress may introduce artifacts in EEG data, potentially influencing the generalization and accuracy of machine learning models.

\section{Methods}
\subsection{Datasets and Preprocessing} \label{sec:Datasets}

\noindent Four public datasets were sourced containing EEG data labeled for acute stress \cite{datasetstress}, schizophrenia \cite{datasetschizophreniatask, datasetschizophreniarest} and healthy controls \cite{datasethealthycontrols}. Table \ref{tab:datasets} shows a summary of each dataset. While both schizophrenia datasets contained data for healthy controls, an additional independent healthy controls dataset was included to ensure a standardized approach to health-screening across control subjects.  \\

\noindent Inclusion and exclusion criteria differed substantially across datasets due to the various mental health conditions present. Of the two schizophrenia datasets included in this study, one \cite{datasetschizophreniarest} confirmed a medication washout period (7 days) prior to EEG recording. For the acute stress dataset \cite{datasetstress}, the inclusion criteria were normal or corrected-to-normal visual acuity, normal color vision, no clinical manifestations of mental or cognitive impairment, verbal or non-verbal learning disabilities. Exclusion criteria were the use of psychoactive medication, drug or alcohol addiction and psychiatric or neurological complaints. The healthy controls dataset \cite{datasethealthycontrols} were labeled as normal by a team of neurologists, after examination of each individual EEG recording.

\begin{table}[h!]
\centering
\caption{\label{tab:datasets}EEG datasets included in this review.}
\resizebox{\textwidth}{!}{
\begin{tabular}{lllllllc}
\hline\hline
\textbf{Study}                        & \textbf{Dataset}              & \textbf{Subjects}   & \textbf{Age}        & \textbf{Gender}            & \textbf{Duration} & \textbf{Channels} & \textbf{Processed}                \\
\hline
\cite{datasetstress}              & Acute Stress (Task)         & 35 Healthy         & 18.6 (Avg) & 9 Male, 26 Female           & 60s      & 23 (500Hz)                  & Y  \\
\hline
\cite{datasetschizophreniatask} & Schizophrenia & 17 + &            & 16 Male, 4 Female & 20 x 6s  & 64 (1200Hz)                & Y  \\
& (Task, Eyes Open) & 23 Healthy & & 15 Male, 7 Female &&& \\
\hline
\cite{datasetschizophreniarest} & Schizophrenia  & 14 + & 28 (Avg)   & 7 Male, 7 Female            & 15m      & 19 (250Hz)                 & Y                     \\
& (Rest, Eyes Closed) & 14 Healthy && 7 Male, 7 Female&&& \\
\hline
\cite{datasethealthycontrols}    & Controls (Rest)     & 100 Healthy       & 23 (Avg)   &                   & 15m      & 32 (200Hz)                & N    \\
\hline
\end{tabular}
}
\end{table}
\FloatBarrier

\noindent All datasets apart from the healthy controls were supplied in preprocessed format (filtered, artifacts removed). To ensure standardization, each dataset was again preprocessed using Python and the MNE-Python \cite{GramfortEtAl2013a} library. Due to the differences in both number of channels and recording frequency of each dataset, these were standardized to the lowest common denominator (200Hz) with only overlapping channels selected (C3, C4, Cz, F3, F4, F7, F8, FP1, FP2, O1, O2, P3, P4, Pz). No temporal channels remained after standardization.\\

\noindent This additional preprocessing included a notch filter (50Hz) to remove any electromagnetic line noise \cite{Newman2020} and a band pass filter was applied between 0.5Hz to 45Hz. Each dataset was re-referenced using average referencing across the remaining channels (14 channels). At 45Hz the down-sampled frequency of 128Hz was within the recommended frequency range (45Hz x 2.5 = 112.5Hz $<$ 200Hz) of interest. \\

\noindent Power Spectrum Density (PSD) was calculated to convert the signals to the frequency domain. The time domain is typically best for viewing and analyzing aperiodic signals \cite{Newman2020}, and this is the basis of much cognitive neuroscience research using event-related potentials (ERPs). However, the frequency domain is better for viewing and analyzing periodic signals. These were generated for alpha, beta, delta, theta and gamma frequencies for frontal, central, parietal and occipital regions, independently.\\

\noindent There are considerable variability across the literature as to the specific frequency range that defines each band \cite{Newson2019}, with significant consequences for results interpretation and study reproducibility. In light of this variability, we opted to define the bands as 0.5Hz to 4Hz for delta, 4Hz to 8Hz for theta, 8Hz to 12Hz for alpha, 12Hz to 30Hz for beta and 30Hz to 45Hz for gamma. For more information and details, please refer to our open-source code and data published alongside this paper on GitHub at https://github.com/xalentis/StressSchizophrenia

\subsection{Machine Learning} \label{sec:ML}

\noindent While SVM has remained the most common classification algorithm for use in schizophrenia prediction using EEG data \cite{Vos2023Review}, tree-based models have shown to be highly accurate, easy to implement, and with good explainability tool support \cite{Arrieta2019, LIME, SHAP}. A number of prior studies have used tree-based models, specifically Random Forest (RF), to predict schizophrenia symptoms with high accuracy (71\% - 99\%) using EEG data \cite{Rahul2024, Ellis2022, Buettner2019, Chu2017}, as well as acute stress on both wearable biomarker \cite{Vos2023, Indikawati2020, Siirtola2020, Nkurikiyeyezu2019} and EEG data \cite{Halim2020, Arsalan2021, Sharif2023, Park2021}. \\

\noindent In this study, XGB \cite{XGBoost} was selected as the machine learning algorithm due to its performance, ease of use and XAI support for both SHAP \cite{SHAP} and Local Interpretable Model-Agnostic Explanations (LIME) \cite{LIME} explainability techniques. Gradient boosting algorithms, such as XGB, outperform RF  in terms of efficiency and predictive accuracy, though they are more prone to overfitting. XGB delivers a highly efficient implementation of the stochastic gradient boosting algorithm \cite{XGBoost} and offers a comprehensive set of model hyperparameters, enabling detailed control over the training process.\\

\noindent Six experiments were designed and run to test a number of key scenarios as follow:

\begin{enumerate}
  \item Test model generalization of schizophrenia task vs. healthy controls (dataset of \cite{datasetschizophreniatask}), and schizophrenia at rest vs. healthy controls (dataset of \cite{datasetschizophreniarest}), independently.
  \item Test model generalization within a multi-class model when both schizophrenia task and rest datasets including their respective healthy controls are combined within the same training dataset to classify healthy, schizophrenia at rest, and schizophrenia when performing a task.
  \item Replace healthy-controls of both schizophrenia datasets with external healthy controls (dataset of \cite{datasethealthycontrols}) and repeat multi-class model experiment.
  \item Identify and exclude all subjects with high acute stress from experiment 3 and repeat the experiment.
  \item Reproduce experiment 4 while applying acute stress correction to  stressed subjects previously excluded.
  \item Validate experiment 5 using a larger pool of external healthy controls identified as non-stressed.
\end{enumerate}

\noindent The first experiment was designed to validate the reproducibility of the high predictive accuracy previously reported in the systematic review by Rahul \emph{et al.} \cite{Rahul2024}. In this preliminary experiment, two XGB models were developed and trained on two datasets: schizophrenia at rest with corresponding healthy controls, and schizophrenia during task performance with corresponding healthy controls. Each model was trained independently and model validation was conducted using leave-one-subject-out (LOSO) validation. Prior to model training, a hyper-parameter search was conducted using 4-fold cross-validation to identify the optimal parameters for the XGB algorithm. \\

\noindent Models were trained for 2,000 rounds with early-stopping set to 10 rounds, thereby ensuring that training stops when the evaluation metric fails to improve after 10 rounds of training, to prevent over-fitting. \\

\noindent In the second experiment, we combined both schizophrenia datasets, designating healthy controls as a single class (class 0) and separating schizophrenia subjects into two distinct classes: at rest (class 1) and during task performance (class 2). The objective was to assess the capability of a machine learning model to differentiate between schizophrenia subjects at rest and those engaged in task performance, while accurately classifying non-schizophrenia healthy controls. Previous studies employing multi-class predictive models on EEG data for mental health classification have demonstrated promising results \cite{Emre2023, Park2021}.\\

\noindent In consideration of the fact that both schizophrenia datasets included their own sets of healthy control EEG data, we conducted the third experiment. In this experiment, the healthy controls within the schizophrenia datasets \cite{datasetschizophreniarest,datasetschizophreniatask} were replaced by an independent, large, and rigorously health-screened dataset \cite{datasethealthycontrols} of subjects confirmed to be healthy, with no identifiable or underlying mental health conditions. This approach was based on the assumption that a robust machine learning model should generalize effectively to unseen data and, crucially, maintain a very low false positive rate during prediction and validation to ensure reliability in clinical settings.\\

\subsection{Acute Stress Artifacts} \label{sec:StressArtifact}

\noindent Experiments 4 and 5 were designed to investigate the potential impact of acute stress during EEG recording on schizophrenia predictive models. For this, a third model was introduced, specifically trained to predict acute stress in EEG data. This model utilized the EEG dataset labeled for acute stress during task performance \cite{datasetstress} and achieved 100\% accuracy using LOSO validation. Our objective was to use this model to identify subjects within the schizophrenia \cite{datasetschizophreniarest,datasetschizophreniatask} and healthy controls datasets \cite{datasethealthycontrols} predicted to be highly stressed (probability $>50\%$). This could also help assess the stress impact on the two multi-class models' performance developed in experiments 2 and 3.\\

\noindent For experiment 5, we considered stress as a potential artifact in schizophrenia EEG data. Our focus was on ensuring only low-stress healthy controls were included in the model training data by potentially filtering this condition out from the frequency bands, as is normally done for ocular and muscular artifacts during EEG data preprocessing. Our approach consisted of extracting stress artifacts by calculating the difference in frequency band power between stressed subjects and non-stressed healthy subjects, and then applying that coefficient to our healthy controls dataset as shown below:\\

\centerline{$AFB = FB - (\overline{FB_s} - \overline{FB_n})$}
Where:
\begin{itemize}
\item $AFB$ is the adjusted frequency band with stress artifacts removed
\item $FB$ is the targeted frequency band to be adjusted
\item $\overline{FB_s}$ is the mean of the frequency band for stressed data
\item $\overline{FB_n}$ is the mean of the frequency band for non-stressed, healthy data
\end{itemize}

\noindent Once the frequency bands of the target training data features were adjusted using the described process, model training and validation were again performed to test whether this adjustment had a positive effect on model generalization. \\

\noindent Finally, in experiment 6 new controls were sampled from a larger pool of the external healthy controls dataset \cite{datasethealthycontrols}. 100 subjects were randomly selected and screened for acute stress using the stress prediction model previously built. Of these, 29 low-stress subjects were selected to form a balanced dataset when combined with the two schizophrenia datasets.\\

\subsection{Explainable A.I. (XAI)} \label{sec:XAI}
\noindent XAI aims to automatically generate a symbolic, human-understandable model from the non-symbolic, statistical machine-learned model \cite{DeBruijn2022}. This human-understandable model is crucial for building trust and validating model performance, while potentially assisting in the understanding of complex data, thereby collectively contributing to the effective and ethical integration of machine learning into  healthcare.\\

\noindent SHAP explainability, when combined with XGB, provides a robust framework for generating an accurate and trustworthy explanation of the predictions made through identifying and ranking important input features \cite{SHAP}. Both the accuracy and explainability of machine learning algorithms are critical in health care, and can further aid diagnosis in cases where patients exhibit similar symptoms, to discern the distinct features that differentiate one diagnosis from another \cite{Hassija2023}.\\ 

\noindent In this study, XAI techniques were employed to address two important questions:

\begin{enumerate}
\item Examine the features identified as significant by the model and assess their correlation with established clinical findings. Identifying features that are both crucial to the model and previously recognized in clinical research can enhance the credibility of machine learning applications in mental health.
\item Identify any feature overlap between schizophrenia at rest and during task performance, to gain insights into the optimal use of predictive models for schizophrenia diagnosis.
\end{enumerate}

\section{Results and Discussion}

\noindent Table \ref{tab:results} details the results obtained for each experiment as described in section \ref{sec:ML}. For experiment 1a, the schizophrenia at rest dataset with its provided healthy controls were used as training/validation input, while experiment 1b utilized the schizophrenia at task dataset with its provided healthy controls. Both experiments utilized a binary classification XGB model with cross-entropy loss, which is based on the Softmax activation function \cite{Belagatti2024}. Experiments 2 to 6 combined the three classes (schizophrenia at rest, schizophrenia at task, and healthy controls) into a multi-class model. Performance was evaluated using LOSO validation with AUC as validation metric. \\

\begin{table}[h!]
\centering
\caption{\label{tab:results}Results of experiments 1-6 explained in Section \ref{sec:ML}.}
\resizebox{\textwidth}{!}{
\begin{tabular}{lcccccl}
\hline\hline
\textbf{Experiment} & \textbf{Controls AUC} & \textbf{Rest AUC} & \textbf{Task AUC}  & \textbf{Data}                               \\
\hline
\rowcolor[rgb]{0.753,0.753,0.753} 1a          & 64\%              & 79\%  &   -       & Original \cite{datasetschizophreniarest}                           \\
1b          & 96\%              &     -     & 82\%  & Original \cite{datasetschizophreniatask}                            \\
\rowcolor[rgb]{0.753,0.753,0.753} 2          & 62\%              & 36\%   & 29\%   & Original \cite{datasetschizophreniarest, datasetschizophreniatask}                            \\
3          & 97\%              & 79\%  & 82\%  & External HC \cite{datasetschizophreniarest, datasetschizophreniatask, datasethealthycontrols}                        \\
\rowcolor[rgb]{0.753,0.753,0.753} 4          & 89\%              & 85\%  & 94\%  & External HC, Low Stress \cite{datasetschizophreniarest, datasetschizophreniatask, datasethealthycontrols}            \\
5          & 97\%             & 86\% & 94\%  & External HC, Adjusted \cite{datasetschizophreniarest, datasetschizophreniatask, datasethealthycontrols}  \\
\rowcolor[rgb]{0.753,0.753,0.753} 6          & 100\%             & 100\% & 100\%  & External HC, Low Stress \cite{datasetschizophreniarest, datasetschizophreniatask, datasethealthycontrols} \\
\hline
\end{tabular}
}
\end{table}

\begin{figure}[h!]
\centering
\includegraphics[width=\textwidth]{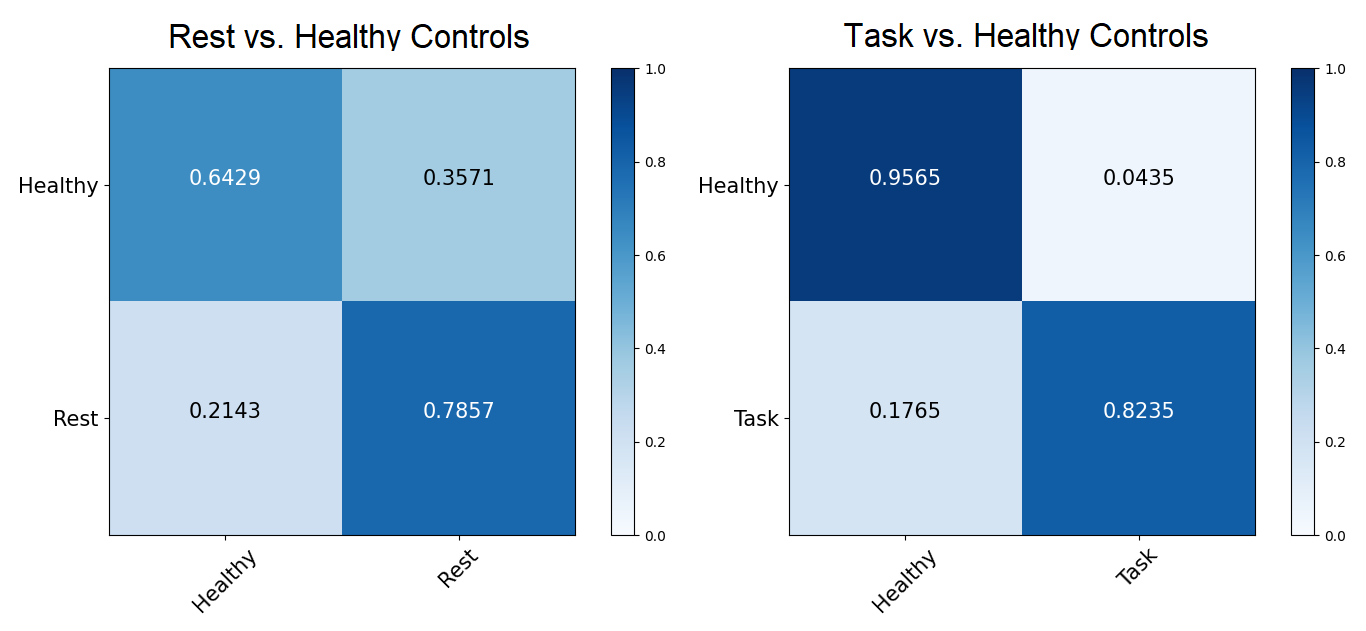}
\caption{\label{fig:figure2}Confusion matrix for experiment 1.}
\end{figure}

\noindent For experiment 1a the AUC score was relatively low compared to those reported by Rahul \emph{et al.} \cite{Rahul2024} in their systematic review of schizophrenia machine learning studies. However, both schizophrenia datasets used in this study contain a relatively small number of subjects. Figure \ref{fig:figure2} details the confusion matrix for both experiments 1a and 1b. \\

\FloatBarrier

\begin{figure}[h!]
\centering
\includegraphics[width=\textwidth]{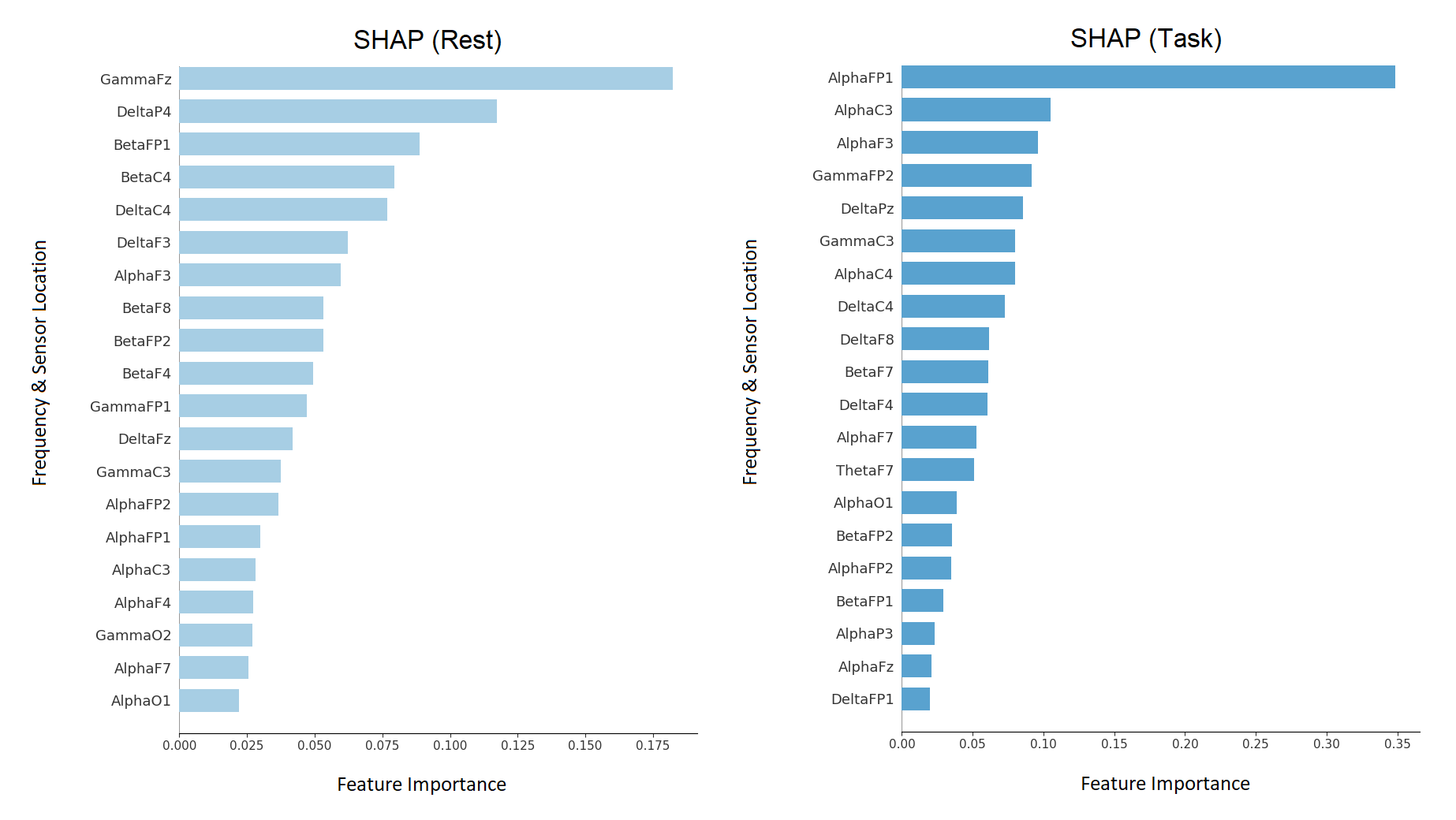}
\caption{\label{fig:figure3}SHAP feature importance for experiment 1.}
\end{figure}

\noindent Figure \ref{fig:figure3} provides the feature importance plots produced by the SHAP explainability technique for both resting and at-task schizophrenia datasets. For schizophrenia at rest, the most important regions are frontal and parietal across the gamma, delta and beta bands, correlating with a number of prior studies \cite{Uhlhaas2010, Newson2019, PerellonAlfonso2023}. For schizophrenia during task performance, alpha, gamma \cite{Uhlhaas2010}, and delta bands were considered the most important features across the frontal \cite{Adamek2022}, central and parietal regions, again consistent with prior studies. These findings show that applying SHAP to machine learning models trained on EEG data is consistent with prior literature and more traditional EEG analysis methods, such as topographic plots of frequency band power. In this context, SHAP explainability provides a further level of model validation beyond test accuracy scores.\\

\FloatBarrier

\begin{figure}[h!]
\centering
\includegraphics[width=\textwidth]{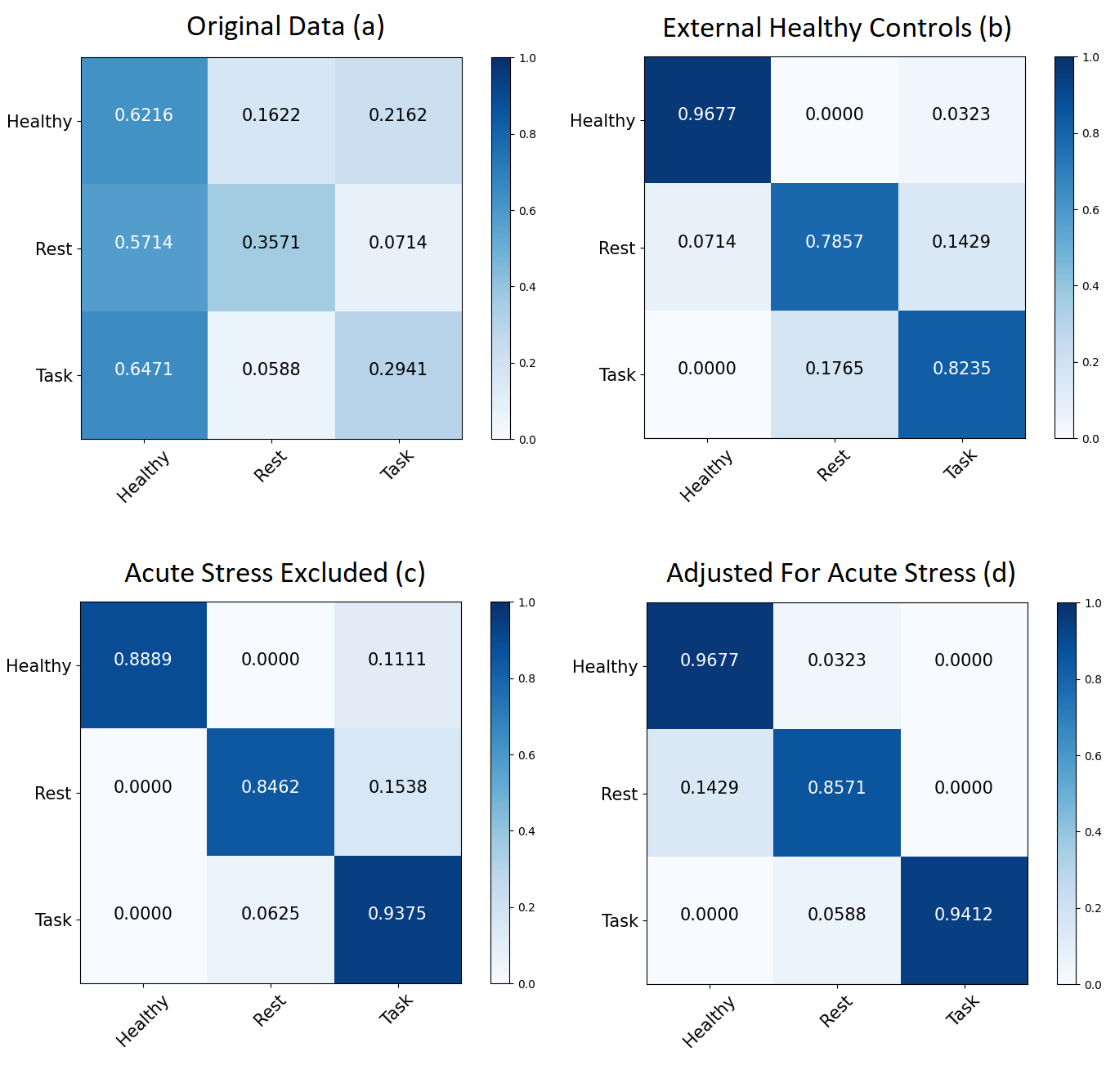}
\caption{\label{fig:figure4}Confusion matrices for experiments 2, 3, 4 and 5 with LOSO validation.}
\end{figure}

\noindent Experiment 2 combined both rest and task datasets into a multi-class dataset where the healthy controls from both datasets were assigned the same class label (0), with schizophrenia at rest and at task assigned classes 1 and 2, respectively. The class-balancing parameter in XGB was set to "weighted" to compensate for the subject count difference between the classes (23+17 in the schizophrenia at task class, and 14+14 for the schizophrenia at rest class). Here the prediction for schizophrenia cases completely failed during the LOSO validation test (Figure \ref{fig:figure4}a), irrespective of class balancing. For the combined healthy controls, predictive accuracy was slightly better, indicating that the model was able to separate healthy subjects from those with schizophrenia while having difficulty separating schizophrenia at rest from schizophrenia at task. This separation is important, considering schizophrenia often affects cognitive functions such as attention, working memory, and executive function, and assessing these areas while the patient is performing a task can provide more detailed and accurate information about the nature and extent of these deficits, compared to when the patient is at rest \cite{Kahn2013}.\\

\FloatBarrier

\noindent To test the effect of health screening and potential underlying conditions including high acute stress during EEG recording \cite{Barros2021}, experiment 2 was repeated, this time using an independent, well-screened external healthy controls dataset. In the resulting experiment 3, healthy controls were removed from the two schizophrenia datasets and replaced with 31 randomly selected healthy controls from the large open NMT Scalp EEG Dataset \cite{datasethealthycontrols}. The results from this experiment were substantially better, as shown in Table \ref{tab:results} and Figure \ref{fig:figure4}b. \\

\noindent Considering the known interaction between schizophrenia and stress \cite{Norman1993, MyinGermeys2001, Vanhollebeke2022, Gomes2016, Vertinski2017, GispendeWied2000, Betensky2008} an additional experiment (4) was performed to understand the improved results from experiment 3 when external healthy controls were utilized. First, an EEG dataset labeled for acute stress was sourced \cite{datasetstress} and a stress prediction model was built using XGB to classify EEG data between healthy, non-stressed subjects and healthy, stressed subjects. The resulting stress prediction model achieved 100\% accuracy using LOSO validation. This model was applied to predict stress on each subject contained in the external healthy controls dataset, and both schizophrenia datasets. \\

\begin{figure}[!h]
\centering
\includegraphics[width=\textwidth]{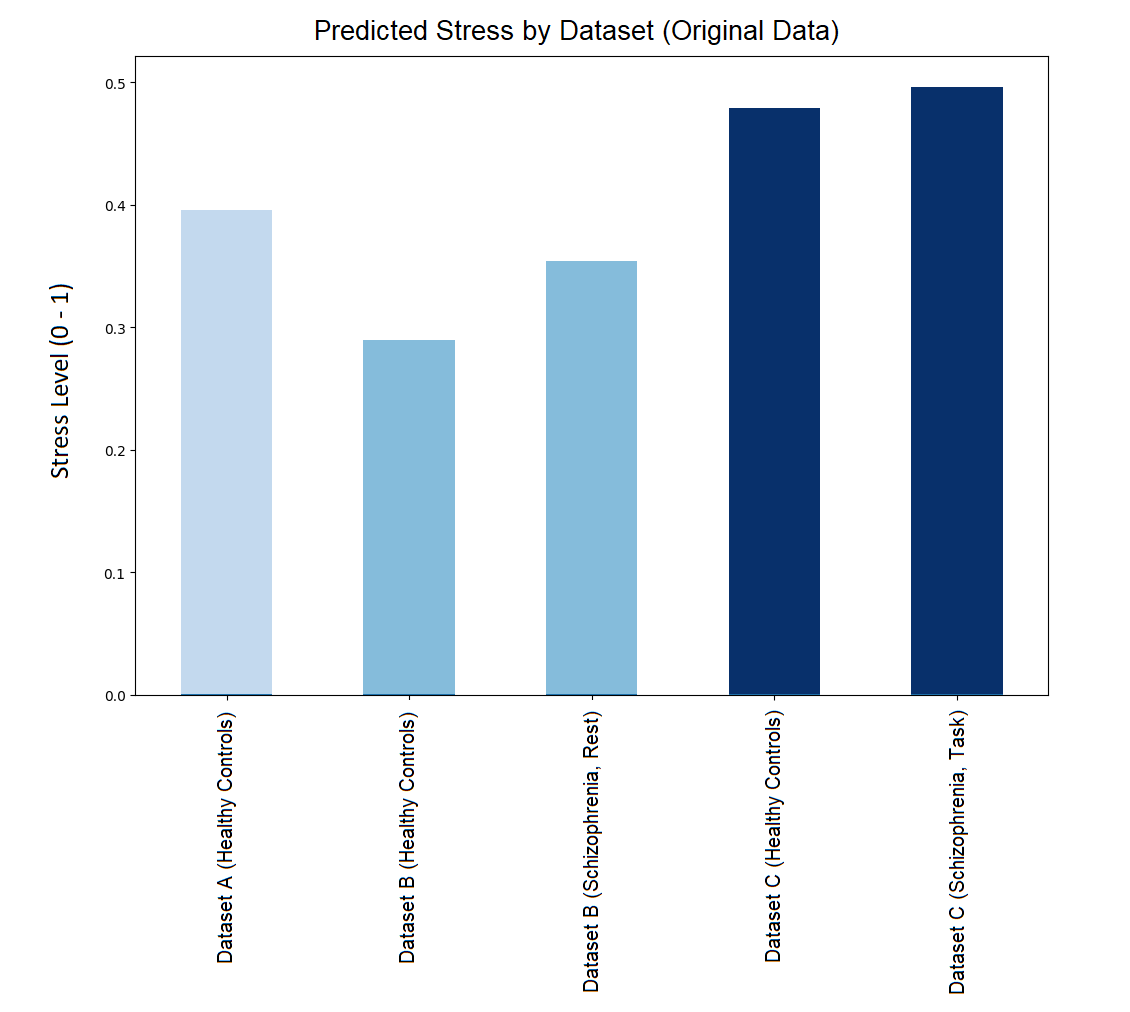}
\caption{\label{fig:figure5}Acute stress prediction on EEG datasets included in this study.}
\end{figure}

\noindent Figure \ref{fig:figure5} shows the average level (mean) of acute stress predicted per dataset, with healthy controls separated from the schizophrenia datasets. In this figure, Dataset A is the external healthy controls dataset from \cite{datasethealthycontrols}, Dataset B is \cite{datasetschizophreniarest}, and Dataset C is \cite{datasetschizophreniatask}. The figure indicates a substantial difference in stress between the healthy controls of both schizophrenia datasets. 

Thirteen healthy controls were predicted with stress probability exceeding the 0.5 threshold and subsequently removed from the external healthy controls dataset A. Additionally, one subject from each schizophrenia dataset with a stress level above the 0.5 threshold was identified and excluded. \\

\FloatBarrier

\noindent Experiment 4 was run as a repeat of experiment 3, using this new dataset containing only low-stress subjects. Results shown in Fig. \ref{fig:figure4}c demonstrate a slight decrease in predictive accuracy for the healthy controls, likely due to the loss of a large number of healthy controls from the dataset, but a substantial improvement for both rest and at-task schizophrenia subjects, with a combined increase in  accuracy score of 18\%. This result confirms an important aspect when utilizing three independent datasets for a 3-class model, in that the model was learning to separate the conditions, as indicated by the drop in accuracy for the healthy controls data, rather than merely learning to classify the actual datasets.\\ 

\begin{figure}[h!]
\centering
\includegraphics[width=\textwidth]{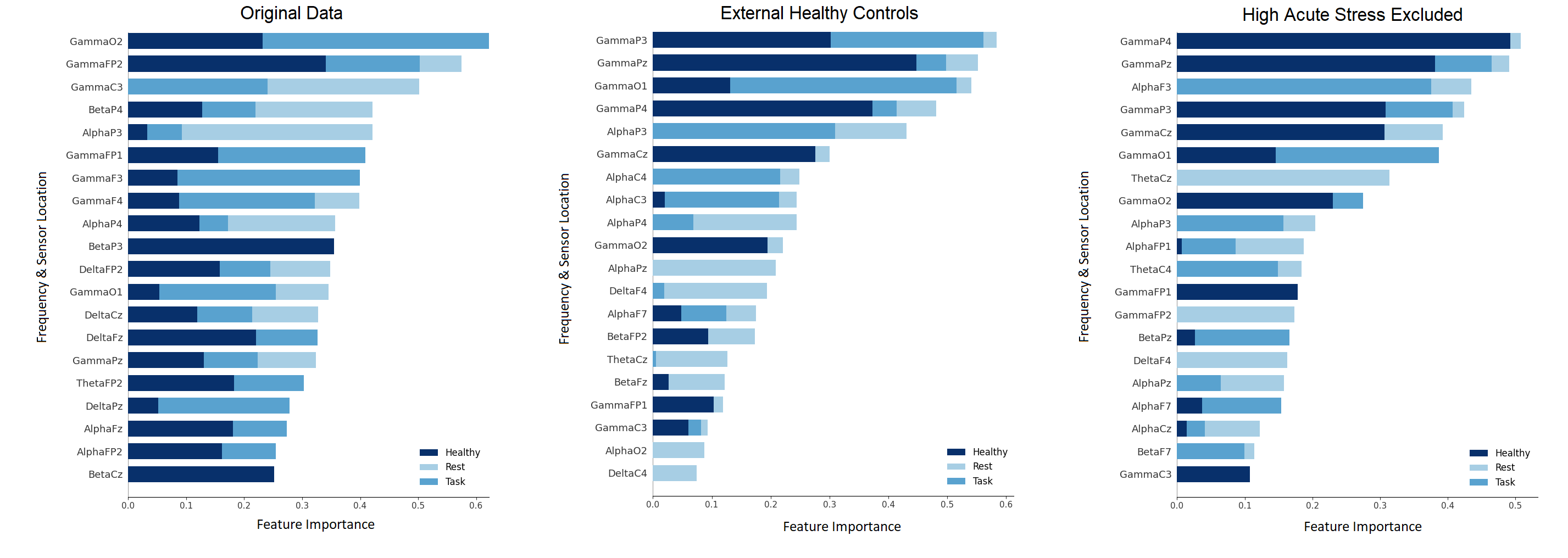}
\caption{\label{fig:figure6}SHAP feature importance for experiments 2, 3, and 4.}
\end{figure}

\noindent SHAP analysis for experiments 2, 3 and 4 are shown in Figure \ref{fig:figure6}, with the gamma band becoming increasingly important \cite{Uhlhaas2010} across all brain regions as stressed subjects are removed from the training data. SHAP analysis further shows increased feature stabilization towards the gamma band when high-stress subjects are excluded, compared to those of the original dataset. We note that for health controls in particular, the gamma band in the parietal region is the most important feature, while gamma in the occipital region along with the alpha band in the frontal region considered to be important for classifying schizophrenia at task subjects. \\

\FloatBarrier

\noindent To address the class imbalance that resulted from removing training data subjects with stress predicted above the 0.5 threshold in experiment 4, a stress coefficient was calculated as described in section \ref{sec:StressArtifact}, and applied to the previously excluded subjects (13 healthy controls, and one from each schizophrenia dataset). Results using this approach with the balanced and stress-adjusted training data are detailed in Table \ref{tab:results}, achieving improved accuracy scores for all three classes, with a significant increase for healthy controls (Figure \ref{fig:figure4}d).\\

\begin{figure}[h!]
\centering
\includegraphics[width=\textwidth]{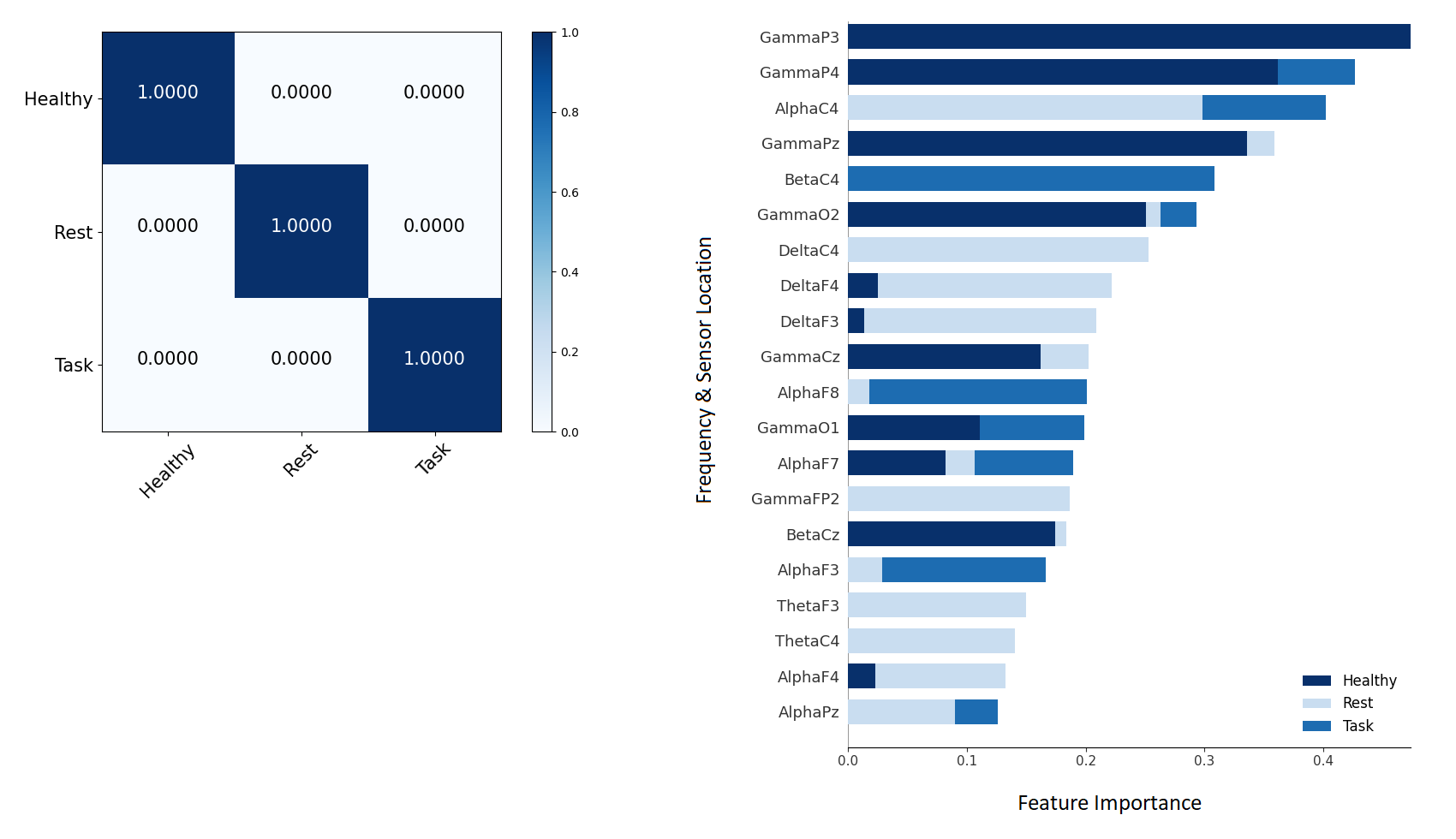}
\caption{\label{fig:figure7}Confusion matrix and SHAP feature importance across balanced datasets after high stress subject exclusion.}
\end{figure}

\noindent To verify this approach to stress artifact removal through frequency band adjustment, a larger pool of 100 subjects was sampled from the healthy controls dataset, and the stress prediction model was again applied in Experiment 6. 29 Subjects from this pool were identified where the probability of stress was predicted below the 0.5 threshold and combined with the remaining 29 schizophrenia subjects (combining the two schizophrenia datasets with 31 subjects and removing the two subjects with high stress). Experimentation was performed on this balanced dataset with the resulting correlation matrix and SHAP analysis depicted in Figure \ref{fig:figure7}. All three classes showed perfect predictive accuracy, with SHAP analysis again showing the gamma band to be the most significant, followed by alpha and beta.

\FloatBarrier

\noindent Importantly, these results confirm the potential impact of acute stress on healthy controls when used as machine learning training data to predict schizophrenia, and the need to either predict for, and remove highly stressed subjects from healthy controls during training, or perform a compensating adjustment to the training data to account for the effect of acute stress on EEG data. \\

\noindent While filtering and artifact removal were pre-applied to the datasets utilized in this study (Table \ref{tab:datasets}), we again applied Independent Component Analysis (ICA) using the MNE-ICALabel \cite{Li2022} library to test whether ICA could assist in identifying acute stress artifacts. We utilized the stress prediction model previously built to predict stress on each subject before, and after applying ICA (10 components) on the schizophrenia at rest dataset. Of the 28 subjects contained within this dataset, 4 schizophrenia subjects and 1 healthy control were predicted with high stress (threshold $>$ 0.4). Once ICA was applied, only a single high-stress schizophrenia subject was sufficiently filtered to reduce stress to a lower threshold. Our findings, therefore, suggest that applying ICA does not provide any benefit in identifying or removing acute stress artifacts from EEG data. \\

\section{Conclusion and Future Work}

\noindent Trustworthiness and robustness are key requirements for the adoption of machine learning techniques for diagnosis and assistance in treating mental health conditions. While robustness can be statistically and mathematically proven, trust requires understanding and verification. XAI techniques can play a crucial role in building this trust by allowing expert clinicians to validate the predictions from machine learning models based on a large body of existing peer-reviewed research literature. \\

\noindent In this study we built a multi-class machine learning model capable of classifying healthy controls from subjects with schizophrenia, at rest and while performing a visual task. We verified the model using SHAP explainability techniques and related those findings to prior literature on schizophrenia. Using the XAI techniques described, we were able to identify overlap between features and proceeded by considering acute stress as a potential effect \cite{Barros2021} on the EEG training data. We proposed and applied a novel technique to test for this effect by removing the stress artifacts from the data, which significantly improved modeling results, ultimately allowing us to select a more reliable set of healthy controls for machine learning training data. Importantly, our findings suggest that proper selection of healthy controls when training machine learning models for prediction on physiological data can play a crucial role in model generalization and robustness. \\

\noindent Given that the methods in this study were applied to a limited amount of publicly available EEG datasets, future research could expand on this work by incorporating explainable AI (XAI) techniques in machine learning studies focused on mental health. This would allow for a deeper examination of the potential effects of comorbidity and patient emotional states on EEG data during session recordings. To support further research in this area, the full source code used in this study is available on GitHub at https://github.com/xalentis/StressSchizophrenia

 \bibliographystyle{elsarticle-num} 
 \bibliography{cas-refs}





\end{document}